\documentclass[twocolumn,showpacs,prl,amsmath,amssymb, superscriptaddress]{revtex4}

\usepackage{graphicx}
\usepackage{amsmath}
\usepackage{amssymb}
\usepackage{bm}

\usepackage{ulem}

\DeclareMathAlphabet{\mathpzc}{OT1}{pzc}{m}{it}

\usepackage{color}

\def\be{\begin{equation}}
\def\ee{\end{equation}}
\def\bea{\begin{eqnarray}}
\def\eea{\end{eqnarray}}

\begin{document}

\newcount\timehh  \newcount\timemm
\timehh=\time \divide\timehh by 60
\timemm=\time
\count255=\timehh\multiply\count255 by -60 \advance\timemm by \count255

\title{Quantum simulation of multiple-exciton generation in a nanocrystal by
    a single photon}
\author{Wayne M. \surname{Witzel} \email{wwitzel@sandia.gov}}
\affiliation{Naval Research Laboratory, Washington DC 20375 USA}
\affiliation{Sandia National Laboratories, NM 87185 USA}
\author{Andrew Shabaev}
\affiliation{George Mason University, VA 22030 USA}
\author{C. Stephen Hellberg$^1$, Verne L. Jacobs$^1$, and  Alexander L. Efros$^1$}
\date{\today, MEG-shortResubmAE.tex, printing time = \number\timehh\,:\,\ifnum\timemm<10 0\fi \number\timemm}
\begin{abstract}
We have shown theoretically that efficient multiple exciton generation
(MEG) by a single photon can be observed in small nanocrystals (NCs). 
Our quantum simulations that include hundreds of
thousands of exciton and multi-exciton states demonstrate that the 
complex time-dependent dynamics of these states in a closed
electronic system yields a saturated MEG effect on a picosecond
timescale. Including phonon relaxation confirms that 
efficient MEG requires the
exciton--biexciton coupling time to be faster than exciton relaxation time. 
\end{abstract}

\maketitle

Solar light would be an important source of clean and renewable energy
if the  efficiency of inexpensive solar cells could be
increased. Increased efficiency can be achieved through carrier
multiplication: Photo-generated
carriers,
  whose excess energy is greater than the energy gap, can create secondary 
  electron-hole pairs via impact ionization of  the filled band.
  Through this process,  two (or more)  
electron-hole pairs are collected 
from each photon
instead
  of just one.
 This process is very inefficient in 
bulk semiconductors, where impact ionization has a very low probability and carrier thermalization, which always competes with impact ionization, is much faster.
 
As first suggested by Nozik, 
impact ionization may  
effectively
compete with cooling  in NCs, 
due to the enhanced rate of inverse Auger processes for 
carrier multiplication and  the ``phonon bottleneck" suppression of carrier relaxation,
leading to efficient  MEG 
\cite{NozikPhysE02}.
Soon after this publication, 
Schaller and Klimov \cite{SchallerPRL04} 
observed 
ultra-efficient MEG by a single photon in PbSe NCs, using band-edge transient absorption measurements. Later,  efficient MEG  was 
observed 
by many groups using different techniques in NCs of many
semiconductors: PbSe
\cite{EllingsonNanoLett05,TrinhNanoLett08,JiNanoLett09,BawendyPbSe},
PbS \cite{EllingsonNanoLett05,SchallerNanoLett06}, Si
\cite{BeardNanoLett07}, CdSe
\cite{ShallerAPL05,SchallerJPC06,GachetNanoLett09}, InAs \cite{SchallerNanoLett07,PijpersJPC-C07},  and in carbon nanotubes \cite{McQuenScience09}. 
At the same time, some groups were not able
to observe MEG in CdSe \cite{BawendyPRB07} and 
InAs \cite{PijpersJPC-C08,BonnNanoLett08} NCs and found
that the efficiency of MEG measured in PbSe NCs \cite{BawendyPbSe} was 
appreciably 
smaller than that reported earlier. 
The diverse experimental data on the MEG efficiency are 
now converging
to more modest values for PbS and PbSe NCs \cite{McGuireACR08}, 
but MEG in NCs has been shown to be significantly more efficient than impact ionization  in bulk semiconductors \cite{NozikNCverBulk,KlimovNCverBulk}. 
           
Previous attempts at a theoretical understanding of 
the enhanced MEG provide only estimations of the MEG 
efficiency observed in NCs
\cite{EllingsonNanoLett05,Klimov-Agranovich,Lanoo-Allan,Zunger,ShabaevNano,Presto,Rebane}.
A self-consistent theory
  of this phenomena requires a currently non-existent 
theoretical description of 
both the relaxation mechanisms for and couplings between
the highly excited exciton and multi-exciton 
states in NCs \cite{ShabaevNano}. 
To explain the high
  efficiency of MEG in NCs, the coherent superposition 
model ,
based on the strong quasi-resonant coupling between exciton and multi-exciton
  states in a NC, was
  proposed \cite{EllingsonNanoLett05,ShabaevNano}.
Non-coherent models 
for
efficient MEG in
  NCs \cite{Klimov-Agranovich,Lanoo-Allan,Zunger,Presto} are based on
the important 
observation that the density of 
biexiton states
is significantly larger than the density of 
exciton states at the same energy
\cite{Klimov-Agranovich}, and the density of trion states is
particularly important for efficient MEG~\cite{Rebane}.
The calculations of MEG efficiency in all of these non-coherent models 
are based on
Fermi's Golden Rule, which requires the final 
biexciton state to decay much faster than the rate of the exciton-biexciton transition, an assumption 
that has not been
justified either experimentally or theoretically.

In this letter we unify both approaches and consider a 
single-photon
excitation  
coherently coupled with 
multi-exciton-states in a NC within a full quantum-state
  evolution approach.  
The time-dependent dynamics of our modeled systems is described using a large 
multiple-configuration basis representation of the many-electron Hamiltonian, 
including energy non-conserving exciton and biexciton decay channels.
The calculations show that even in a closed, energy-conserving electronic system, the excitation becomes  
predominantly multi-excitonic
on a picosecond time scale.  
The initial single-photon excitation
is dispersed into the dense multi-exciton state space of the NC.

We consider a single-photon excitation of a spherical PbSe NC,
where efficient MEG has been reported
\cite{SchallerPRL04,EllingsonNanoLett05,TrinhNanoLett08,JiNanoLett09,BawendyPbSe}.
The energy spectrum of electrons and holes at the band edges of four equivalent L 
valleys of bulk PbSe is described  by a four band ${\bf k} \cdot {\bf p}$ model 
\cite{MitchellPR66}. In PbSe NCs
each electron, $n^eL^j_m$, and hole, $n^hL^j_m$, state within  
the four-band effective-mass model \cite{KangWise},
 is characterized by the  spatial angular momentum $L=0(S),1(P),2(D),3(F),...$, the total angular momentum $j=L\pm1/2$, which is the sum of the angular momentum and the spin, the projection of the total momentum $m=\pm 1/2, \pm 3/2,...\pm j$,
and the spatial inversion parity $\pi$. The energy levels of the same
symmetry were calculated using energy band parameters from
Ref.~\onlinecite{EllingsonNanoLett05} and labeled by a level number, $n^{e,h}$.
The four envelope components, $F_i({\bm r})$, ($i=1,2,3,4$) of the corresponding single-particle wavefunctions can be expressed in the form:
\be
\label{envelopeForm}
F_i({\bm r}) = C^{L, m, \pi}_i  f^n_{\tilde{L}_i^{\pi}}(r)
Y_{\tilde{L}_i^{\pi}}^{\tilde{m}_i^{\pi}}(\theta, \phi),
\ee
where $C^{L, m, \pi}_i$ are constants, $f^n_{l}(r)$ are specific
 radial functions expressed in terms of
the spherical Bessel functions,
 $\tilde{L}_i^{\pi} = L~\mbox{or}~L+1$ and $\tilde{m}_i^{\pi} = m \pm 1/2$
(depending upon the values of $i$ and $\pi$), and
$Y_{\tilde{L}_i}^{\tilde{m}_i}(\theta, \phi)$ are the 
spherical harmonics.

\begin{figure}
\includegraphics[width=3.5in]{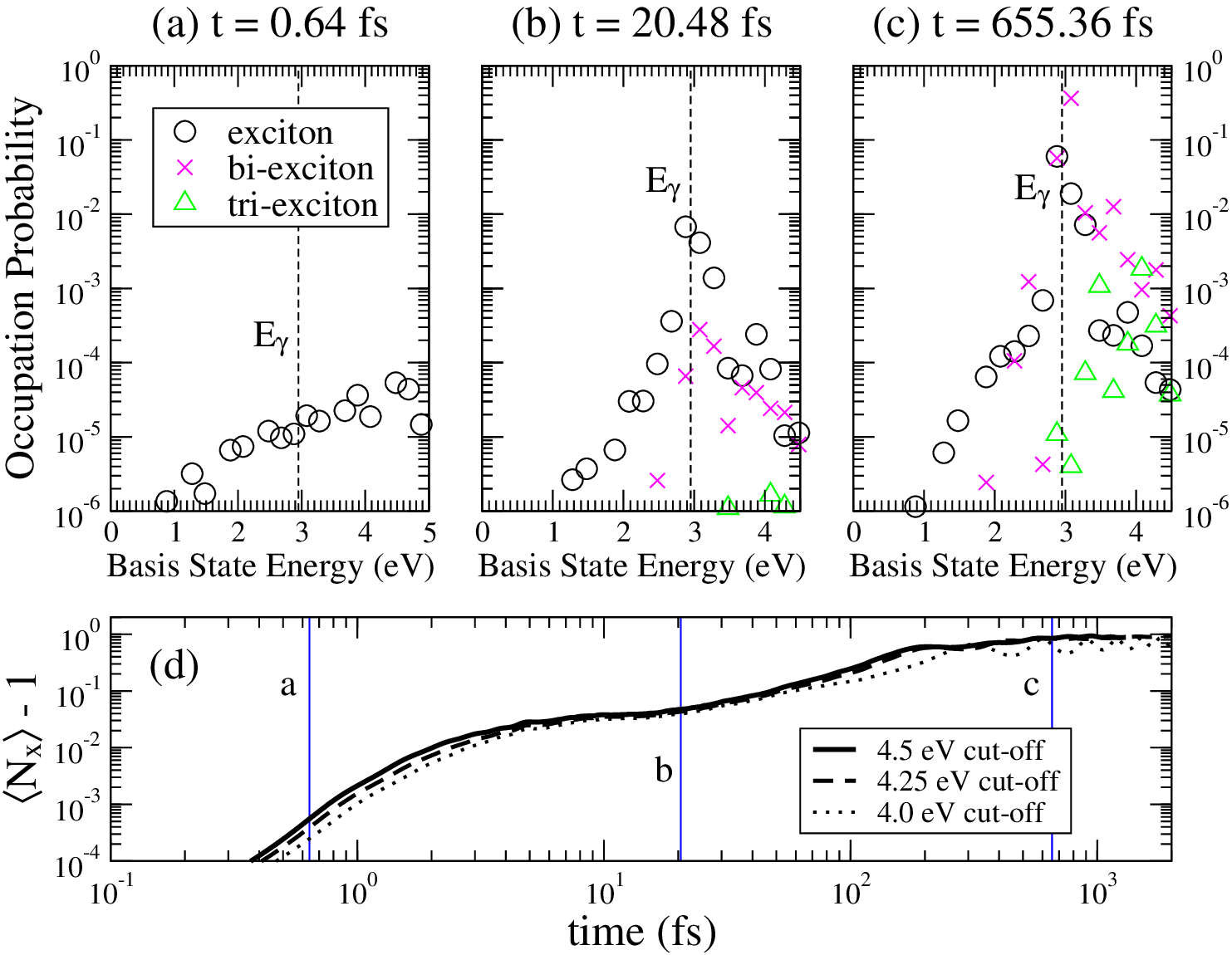}
\caption{
\label{energyProbTimeSeq}
Time dependent evolution of
the $2^eP^{1/2}_{1/2}-2^hP^{1/2}_{1/2}$ excitation created in the
2\,nm radius PbSe NC with $\kappa_s = \kappa_g = 5$ by a single 
$E_\gamma = 2.95~\mbox{eV}$ 
photon
calculated for the
closed-system. 
The upper panels (a,b,c) show  the occupation probablities  of  exciton, bi-exciton and tri-exciton states over $0.2~\mbox{eV}$
wide
energy bins 
at three times.
Panel (d) demonstrates convergence of the time dependence of the relative number of excitons $\langle N_{x}(t) \rangle - 1$, defined in Eq.~\eqref{Nexpectation},
achieved by adjusting the energy cut-off. 
The vertical 
lines
a, b and c show the 
times corresponding to the three upper panels.} 
\end{figure}

The Coulomb coupling of an electron 
or hole 
with multi-electron-hole excitations at the same energy, 
which are disallowed in the bulk due to momentum conservation,
plays an important role for MEG in NCs \cite{ShabaevNano}. 
We generate a Hamiltonian matrix in
the basis of many-electron states that are Slater determinants of  
single-particle eigenstates.  
A many-electron state in this approach is a set of occupied
conduction states (electrons) and unoccupied valence states (holes)
with an arbitrary but consistent sign convention for the 
Slater-determinant permutations. We include many-electron basis states up to some chosen
energy cut-off that we adjust until the results converge (see
Fig.~\ref{energyProbTimeSeq}). 
We modeled the
Coulomb interaction using effective dielectric constants  
for the
NC, 
$\kappa_s$, and surrounding media, $\kappa_g$,  
leading
to the direct Coulomb interaction between electrons at positions 
$\bm r$ and $\bm r'$ in the NC,
$V_C({\bm r}, {\bm r'}) = e^2/(\kappa_s |{\bm r} - {\bm r'}|$),
and the surface mediated Coulomb potential \cite{Jackson}, $V_s({\bm r}, {\bm r'})=e^2a/(\tilde{\kappa}r'|{\bm r} - a^2{\bm r'}/r'^2|)$, where $a$ is the NC radius and $\tilde{\kappa}^{-1}= 2 (\kappa_s - \kappa_g)/\kappa_s (\kappa_s + 2 \kappa_g)$.
We also account for the
  electron and hole interaction with their image potentials
  described by $V_e(r)=V_h(r)=0.5V_s({\bm r}, {\bm r})$.

The diagonal elements of this Hamiltonian matrix 
are taken
to be sums
of the single-particle-state energies (including image potentials)
and the Coulomb interactions
among electrons and holes 
(including the exchange interactions from Fermi permutations).
Off-diagonal elements 
couple basis states
that differ in either two or four sets of 
single-state
occupations via the Coulomb interaction.

Because the single-particle states in the spherical NCs are represented in terms of
spherical harmonics, it is most efficient to express the Coulomb 
potentials $V_C({\bm r}, {\bm r'})$ and $V_s({\bm r}, {\bm
    r'})$   in terms of spherical harmonics \cite{Jackson}:
\begin{eqnarray}
V_C({\bm r}, {\bm r'}) &=& \frac{4 \pi e^2}{\kappa_s} \sum_{l=0}^{\infty} \frac{1}{2 l + 1}
\frac{r_<^l}{r_>^{l+1}} \Upsilon_l(\theta, \phi, \theta', \phi'),
\\
\nonumber
V_S({\bm r}, {\bm r'}) &=& \frac{4 \pi e^2}{a \tilde{\kappa}} 
\sum_{l=0}^{\infty} \frac{1}{2l + 1} \left(\frac{r r'}{a^2}\right)^{l}
\Upsilon_l(\theta, \phi, \theta', \phi'),
\end{eqnarray}
where $\Upsilon_l(\theta, \phi, \theta', \phi') = \sum_m Y_{lm}^*(\theta', \phi') Y_{lm}(\theta, \phi)$,
${\bm r}$ and ${\bm r'}$ are represented by $\{r, \theta,
  \phi\}$  and $\{r', \theta', \phi'\}$ in spherical coordinates, 
  $r_> = \max{(r, r')}$, and $r_< = \min{(r, r')}$.  
Using these expressions, we have performed integrations 
involving the angular components analytically by exploiting the
orthonormality of the spherical harmonics. 

The remaining radial parts of the integrals of 
the Coulomb matrix elements take the form
$\int dr \int dr' f^m_{i}(r) f^n_{j}(r) f^o_{k}(r') f^p_{l}(r') 
[(r_<)^l / (r_>)^{l+1}]$, or 
$\int dr \int dr' f^m_{i}(r) f^n_{j}(r) f^o_{k}(r') f^p_{l}(r') \left(1
/ r r'\right)^{l+1},$
where $f^{m}_{i}(r)$ are the radial functions defined in
 Eq.~(\ref{envelopeForm}) for the envelope 
eigenfunctions that are expressed via spherical Bessel functions.  
In our computational implementation,
we approximate these Bessel function integrals as needed  by 
Monte-Carlo sampling and store the result for future reference. 
A sufficient number of samples was taken to reach a 
specified level of precision for the matrix
elements \cite{precision}.
At a 4.8 eV cut-off used to simulate the $2~\mbox{nm}$
  radius NCs, the basis contains hundreds of thousands
of many-electron
states, and the Hamiltonian matrix contains tens of millions of non-zero elements.
  
To incorporate the interaction involving a single photon, 
we 
adopt
the standard
effective-mass approximations that the wavelength of the photon is much larger
than the NC size and that the contributions of the
Bloch functions are dominant in the integral over the 
electron momentum operator \cite{KangWise}.  
In the PbSe NCs with symmetric conduction and valence bands,
we assume that
the photons are coupled exclusively to symmetric
electron-hole pairs, in which both the electron and the hole 
have the same quantum numbers.  
As a result the
NC-photon interaction
is characterized by a single 
parameter corresponding to the coupling energy. 
Furthermore, our results
are obtained in the weak-coupling 
single-photon limit, 
 where the actual coupling strength is irrelevant provided that it is
 sufficiently small.

Let us begin from a study of the closed system evolution.
Starting from an initial single-photon-state,
electronic states are
excited and evolve through Coulombic interactions.
In each simulated experiment, the photon energy was chosen
to match one
of the optically active exciton states, because the NC absorption
spectra are not affected by the coupling between excitons and
multi-exciton states \cite{ShabaevNano}.  We evolve the
initial state using the Schr\"odinger
equation, $d \lvert \Psi(t) \rangle / dt = -i \hat{H} \lvert \Psi(t)
\rangle / \hbar$, where $\hat{H}$  is our 
Hamiltonian that includes all interactions considered above. 
Using a
differental equation
  solver \cite{CVODE} for integration, we adjusted the precision until we
obtained convergent results.

\begin{figure}
\includegraphics[width=3.5in]{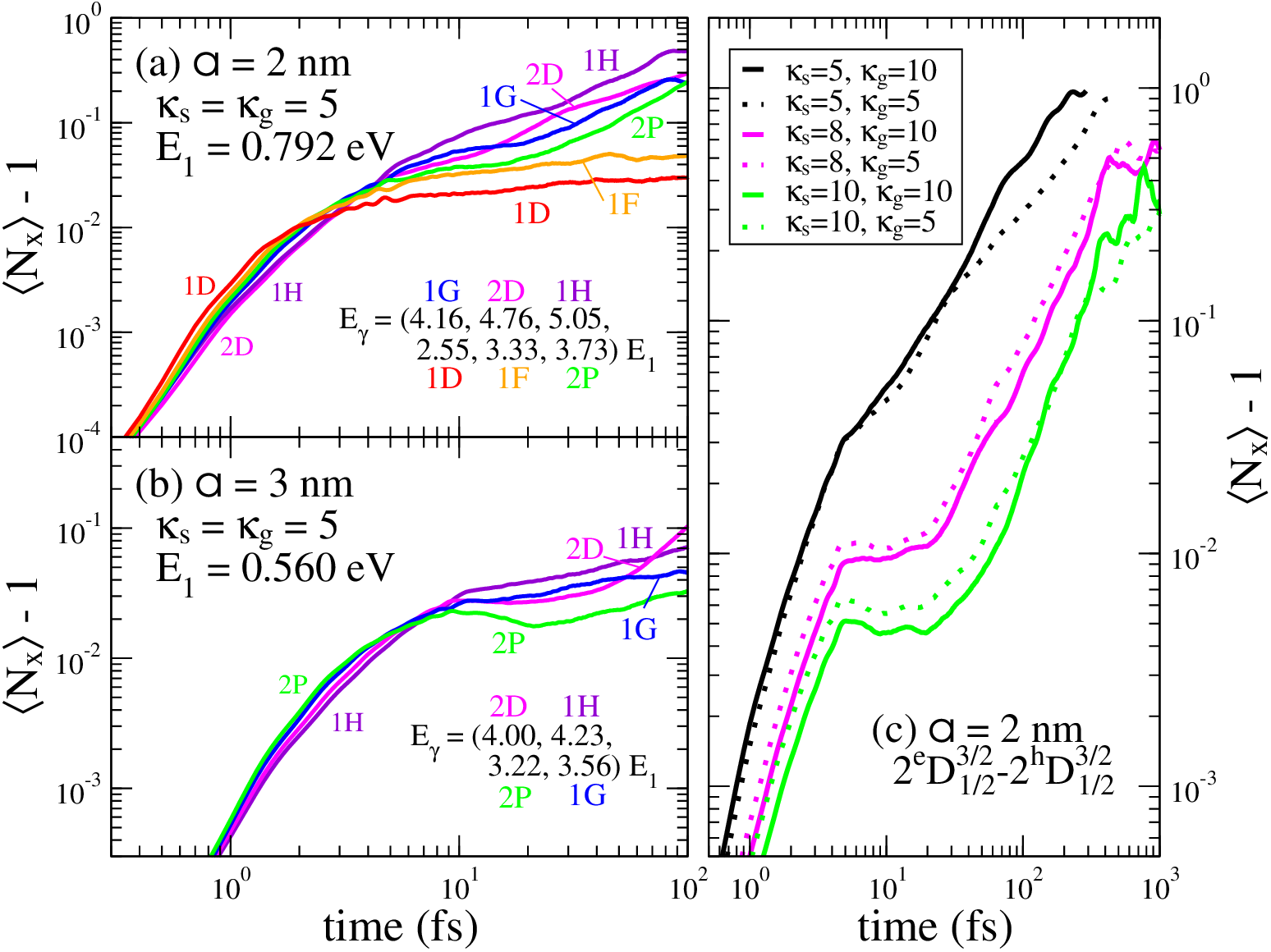}
\caption{
\label{compareDotSizeAndDielectric}
Time dependent evolution of the relative number of excitons 
$\langle N_x (t)\rangle$,
defined in Eq.~\eqref{Nexpectation},
created by a single-photon 
excitation of various optical transitions in PbSe NCs with radii $2$\,nm 
(a) and $3$\,nm (b) 
calculated for the
closed-system 
with $\kappa_s=\kappa_g=5$. 
The optical transitions $2^eP^{1/2}_{1/2}-2^hP^{1/2}_{1/2}$,
$1^eD^{3/2}_{1/2}-1^hD^{3/2}_{1/2}$,
$1^eF^{5/2}_{1/2}-1^hF^{5/2}_{1/2}$,
$2^eP^{1/2}_{1/2}-2^hP^{1/2}_{1/2}$,
$1^eG^{7/2}_{1/2}-1^hG^{7/2}_{1/2}$,
$2^eD^{3/2}_{1/2}-2^hD^{3/2}_{1/2}$,
$1^eH^{9/2}_{1/2}-1^hH^{9/2}_{1/2}$, are abbreviated $1D$, $1F$, $2P$,
$1G$, $2D$, and $1H$ respectively with energies shown in units
of the effective energy gap $E_1(a)$. Panel (c) shows the 
evolution of $\langle N_x(t) \rangle$ created by a single-photon
excitation of the $2^eD_{1/2}^{3/2}-2^hD_{1/2}^{3/2}$  transition in a
2\,nm radius PbSe NC calculated for various combinations of the
dielectric constants of the 
nanocrystal $\kappa_s$ and the surrounding
media $\kappa_g$.}
\end{figure}

Figure~\ref{energyProbTimeSeq} shows the time-dependent evolution of
the $2^eP^{1/2}_{1/2}-2^hP^{1/2}_{1/2}$ excitation created by a single photon   calculated for our
closed-system description. The upper panels (a,b,c) show  the occupation probablities  $\|\langle k \vert \Psi(t)
\rangle \|^2$ of  exciton, bi-exciton and tri-exciton states created
from the photon at three different snapshots in time. 
The initially created excitons are replaced by the
biexcitons in approximately 0.3 ps. 
Formation of
tri-excitons at the excitation energy considered,  $2.95 eV \sim 3.73E_1(a)$ where
$E_1(a)$ is the effective energy gap,  is limited,
however, in our single valley model, 
which allows only two electron-hole pairs to occupy the ground state.  
Figure~\ref{energyProbTimeSeq}(d) shows the time dependence of the expectation value
for the relative number of excitons:
\begin{equation}
\label{Nexpectation}
\langle N_{x}(t) \rangle = \sum_{n=1}^3 n \sum_{k \in {\cal K}_n} \|\langle k \vert \Psi(t)
\rangle \|^2 / \sum_{l>0} \|\langle l \vert \Psi(t)
\rangle \|^2,
\end{equation}
where ${\cal K}_n$ is the set of $n$-exciton basis states.
In the small NC with small dielectric constant considered
in Fig.~\ref{energyProbTimeSeq}, the strong Coulomb
coupling rapidly creates a bi-exciton with $90\%$ probability.

We also investigate how  
$\langle N_{x}(t) \rangle$ depends on the excitation energy and the 
Coulomb
coupling, which in our model is controlled by the NC size and dielectric constants. 
Figure~\ref{compareDotSizeAndDielectric}(a) compares the formation of
multi-excitons from a single photon excitation of different optical
transitions in the $2$\,nm radius NC. The $2^eP^{1/2}_{1/2}-2^hP^{1/2}_{1/2}$ transition appears to provide a
  threshold for efficient MEG generation in PbSe NCs: while lower
  transitions saturate at around $10~\mbox{fs}$ to a low probability
  of biexciton generation, the multi-exciton
   number continues to
  increase when the threshold is reached.
This is because the $2^{e}P$ and $2^{h}P$ states are the first 
that can be strongly coupled with trions of the same
energy \cite{ShabaevNano}.  Increasing the NC size
[Fig. \ref{compareDotSizeAndDielectric}(b)]  and the dielectric constant [Fig. \ref{compareDotSizeAndDielectric}(c)] significantly increases
the time of MEG generation.   Both effects  are 
the direct result of the reduced Coulomb coupling. 
 
Note that
the evolution of $\langle N_{x}(t) \rangle$ does not show strong oscillations
when MEG is efficient, even for a closed system. Oscillations 
are seen
only in the striking 
cross-over behavior  in
Fig.~\ref{compareDotSizeAndDielectric} (a) and (b), which 
is the result of the different coupling strengths and detunings between the multi-exciton states and the exciton created by the resonant light. 
The coherent oscillations connecting the exciton and biexciton states are suppressed due to 
the large number of
interacting biexciton states. 
Once the excitation evolves 
into the dense multi-exciton state space, it does not appreciably revive.
 
Phonon-assisted assisted relaxation of carriers, which was not considered above,
can strongly affect the MEG efficiency~\cite{ShabaevNano}.
The evolution of $\langle
N_{x}(t) \rangle$, including phonon decay, is described by
\begin{figure}
\includegraphics[width=3.5in]{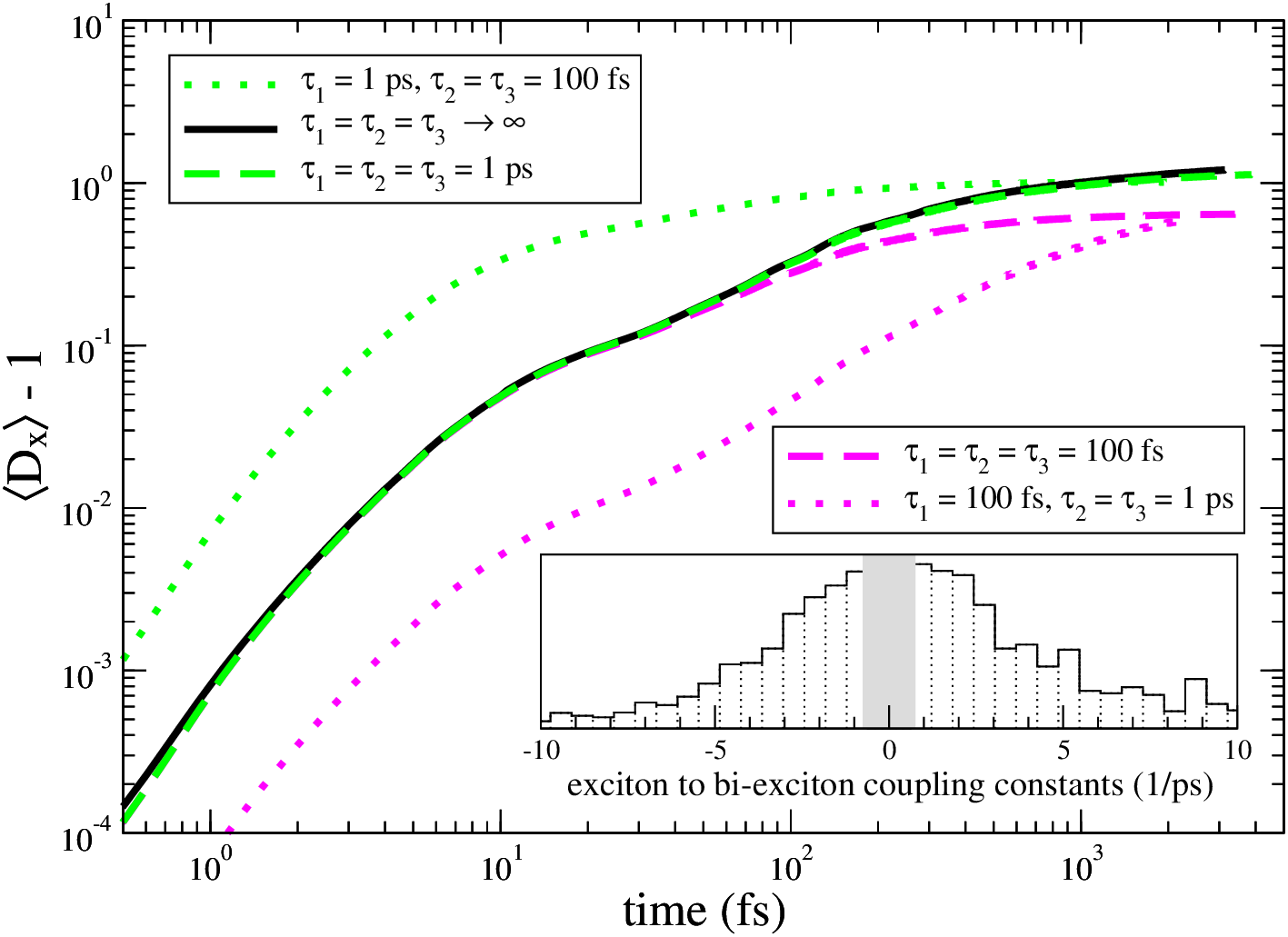}
\caption{\label{decayCompare}
Evolution of the average number of decayed-to-the-ground-state excitons
$\langle D_{x}(t) \rangle$ created by a single-photon excitation  of
the $1^eH^{9/2}_{1/2}-1^hH^{9/2}_{1/2}$ transition of 
a $2~\mbox{nm}$ radius PbSe NC  calculated for various combinations of  
thermalization times.   
The inset
shows the distribution of inverse coupling times connecting optically active exciton states and bi-exciton states, which are equal to the interaction Hamiltonian matrix elements,
divided by $\hbar$. 
The 1/ps scale in the inset allows 
comparison of typical coupling times to thermalization times. 
The gray region of the coupling strengths is unresolved at our precision 
and excluded from the distribution.}
\end{figure}
\begin{equation}
\label{generalizedSchrodinger}
\frac{d}{dt} \lvert \Psi \rangle = -\frac{i}{\hbar} \hat{H} \lvert \Psi \rangle 
- \sum_n \sum_{k \in {\cal K}_n} \frac{1}{2 \tau_n} \langle k \vert \Psi \rangle
\lvert k \rangle,
\end{equation}
where $\tau_n$ is the relaxation time
of the $n$-exciton state that for simplicity is assumed to
  depend only on $n$.
As states decay, they are removed from the Hilbert space and populate
  $D_n(t)$, denoting each ``ground'' $n$-exciton state and governed by
$\dot{D}_n = \sum_{k \in {\cal K}_n} \| \langle k
  \vert \Psi \rangle \|^2 / \tau_n$ with $D_n(0)=0$.

To find the relative number of excitons following decay, we take the
weighted average $\langle D_{x} \rangle = \sum_n n D_n / \sum_m D_m$.  We plot this
average in
Fig.~\ref{decayCompare} for various $\tau_n$ lifetimes.  
The introduction of the long and equal relaxation times
$\tau_1=\tau_2=\tau_3=1$\,ps has little impact upon the time dependent
evolution: the system remains  almost closed. 
Indeed, these times are much longer than an average time of the exciton-biexciton coupling as one can see in the inset of Fig.~\ref{decayCompare}. 
Reducing
the biexciton and trion decay times to $\tau_2=\tau_3=100$\,fs 
increases
the rate of MEG in agreement with Ref. \cite{ShabaevNano}.   
It is
interesting to note that  the saturation value for MEG depends only
upon $\tau_1$. If this time is shorter  than the exciton-biexciton coupling time,
$\sim 300$\,fs, MEG is significantly suppressed.

Our calculations of the dynamics and efficiency of MEG, unfortunately,
cannot be compared  with experimental data measured in PbSe NCs
because they were based on a single-valley model. The lack of a
precise knowledge of inter-valley coupling does not allow us to take this effect into account.  The single-valley approximation should significantly underestimate the density of 
multiple-exciton states. For example, in real PbSe NCs one can have
eight electron-hole pairs at the ground exciton state instead of
only the two ground excitons allowed in our model. The inter-valley coupling significantly increases the density of 
  multi-exciton states; the $k$-exciton state density
    is increased by at least a factor of $4^{2k}$ for PbSe with its $4$ valleys.
For this reason,
  MEG should be even more efficient in a model that includes
  inter-valley coupling.

In summary, our calculations unambiguously demonstrate that highly
efficient MEG can be observed in small
NCs. The effect 
is enhanced by a high density of biexciton states that are strongly
coupled with optically created excitons. 
Fast multi-exciton thermalization accelerates the
formation of multi-excitons in the ground state which should improve
efficiency of extraction of electron hole pairs from NCs.

We thank Andrew Taube for 
important  
scientific advise.  
We
acknowledge financial support from ONR. A.S. acknowledge support from
NIST 70NANB7H6138 Am 001, and Center for Advanced Solar Photophysics,
a DOE Energy Frontier Research Center.

Sandia National Laboratories is a multi-program laboratory 
operated by Sandia Corporation, a 
wholly owned subsidiary of Lockheed Martin Corporation, for the U.S. Department of Energy's National Nuclear
Security Administration under contract DE-AC04-94AL85000.


\begin{thebibliography}{99}
\bibitem{precision}  The level of precision was tested by repeating the calculations with a different random-number-generator seed
and checking for consistency.
\end{thebibliography}
\end{document}